\documentclass[10pt]{iopart}
\usepackage{epsfig}
\usepackage[numbers]{natbib}

\def\lesssim{\mathrel{\hbox{\rlap{\hbox{\lower4pt\hbox{$\sim$}}}\hbox{$<$}}}}
\def\gtrsim{\mathrel{\hbox{\rlap{\hbox{\lower4pt\hbox{$\sim$}}}\hbox{$>$}}}}

\def\lsim{\mathrel{\hbox{\rlap{\hbox{\lower4pt\hbox{$\sim$}}}\hbox{$<$}}}}
\def\gsim{\mathrel{\hbox{\rlap{\hbox{\lower4pt\hbox{$\sim$}}}\hbox{$>$}}}}

\usepackage{ulem}

\newcommand{\K}{\;\mathrm{K}}

\newcommand{\s}{\;\mathrm{s}}
\newcommand{\erg}{\;\mathrm{erg}}

\newcommand{\yr}{\;\mathrm{yr}}
\newcommand{\Myr}{\;\mathrm{Myr}}
\newcommand{\Gyr}{\;\mathrm{Gyr}}

\newcommand{\Gpc}{\;\mathrm{Gpc}}
\newcommand{\Mpc}{\;\mathrm{Mpc}}

\newcommand{\Msol}{\;\mathrm{M}_{\odot}}

\newcommand{\apj}{ApJ}
\newcommand{\apjl}{ApJ}
\newcommand{\apjs}{ApJS}
\newcommand{\aap}{A$\&$A}
\newcommand{\aapr}{A$\&$A}
\newcommand{\araa}{ARAA}

\newcommand{\mnras}{MNRAS}

\newcommand{\physrep}{Physical Reports}
\newcommand{\aj}{AJ}
\newcommand{\nat}{Nature}

\def\beq{\begin{equation} }
\def\eeq{\end{equation} }
\def\spose#1{\hbox to 0pt{#1\hss}}
\def\ltsim{\mathrel{\spose{\lower.5ex\hbox{$\mathchar"218$}}\raise.4ex\hbox{$\mathchar"13C$}}}

\def\spose#1{\hbox to 0pt{#1\hss}}
\def\lta{\mathrel{\spose{\lower 3pt\hbox{$\mathchar"218$}}
        \raise 2.0pt\hbox{$\mathchar"13C$}}}
\def\gta{\mathrel{\spose{\lower 3pt\hbox{$\mathchar"218$}}
        \raise 2.0pt\hbox{$\mathchar"13E$}}}
\def\newblock{\hskip .11em plus .33em minus .07em}

\begin{document}

\title[Growing the First Quasar Black Holes]{Driving the Growth of the Earliest Supermassive Black Holes with Major Mergers of Host Galaxies}

\author{Takamitsu L. Tanaka$^{1}$}

\address{$^1$ Max Planck Institute for Astrophysics, Karl-Schwarzschild-Str.~1, D-85741 Garching, Germany}

\ead{taka@mpa-garching.mpg.de}

\begin{abstract}
The formation mechanism of supermassive black holes (SMBHs) in general,
and of $\sim 10^9\Msol$ SMBHs observed as luminous quasars at redshifts $z> 6$ in particular,
remains an open fundamental question.
The presence of such massive BHs at such early times,
when the Universe was less than a billion years old,
implies that they grew via either super-Eddington accretion, or
nearly uninterrupted gas accretion near the Eddington limit;
the latter, at first glance, is at odds with 
empirical trends at lower redshifts, where quasar episodes
associated with rapid BH growth are rare and brief.
In this work, I examine whether and to what extent
the growth of the $z> 6$ quasar SMBHs can be explained
within the standard quasar paradigm,
in which major mergers of host galaxies trigger
episodes of rapid gas accretion below or near the Eddington limit.
Using a suite of Monte Carlo merger tree simulations of
the assembly histories of the likely hosts of the $z> 6$ quasars,
I investigate
(i) their growth and major merger rates out to $z\sim 40$,
and (ii) how long the feeding episodes induced by host mergers must last
in order to explain the observed $z\gta 6$ quasar population
without super-Eddington accretion.
The halo major merger rate scales roughly as $\propto (1+z)^{5/2}$, with
quasar hosts typically experiencing 
$\gta 10$ major mergers between $15> z > 6$ ($\approx 650\Myr$),
compared to $\sim 1$ for typical massive galaxies at $3>z > 0$  ($\approx 11 \Gyr$).
An example of a viable sub-Eddington SMBH growth model is one where
a host merger triggers feeding for a duration comparable to the halo dynamical time.
These findings suggest that the growth mechanisms of the
earliest quasar SMBHs need not have been drastically different
from their counterparts at lower redshifts.
\end{abstract}

\section{Introduction}
\label{sec:intro}
Observations have established the presence
of a supermassive black hole (SMBH) in the center of
virtually every massive galaxy in the local Universe \cite{KR95}.
There is a large body of circumstantial evidence suggesting
that feedback from SMBHs during luminous accretion
episodes---active galactic nuclei or quasars---plays prominent
roles in galaxy evolution \cite[][and references within]{KormendyHo13}.
Quasar activity also helped to reionize and heat the intergalactic medium
\cite{Madau+04, FanReview06},
which may have influenced the formation and evolution of low-mass galaxies
and their central BHs \cite{Gnedin00, Ripamonti+08, TPH12, TLH13, Jeon+14}.

The origins of these cosmic behemoths 
remain a fundamental unsolved problem
\cite[see reviews by][]{Volonteri10, Haiman13, Natarajan14}.
Particularly puzzling are the SMBHs with masses in excess of $10^9\Msol$
powering luminous quasars at redshifts $z\gta6$ \cite{Fan+01, Fan+03, Willott+03, Willott+10a, Mortlock+11, Venemans+13, DeRosa+13}, less than 
$1 \Gyr$ after the big bang. 
To reach such masses in so short a time, these SMBHs
must have either accreted nearly continuously near
the Eddington limit \cite[e.g.][]{Shapiro05, Pelupessy+07, TH09}
or undergone episodes of super-Eddington accretion \cite{King03, VolRees05, Begelman12, WL12, VolSilk14, Madau+14}---regardless
of whether they formed as the remnants of the
first generation of stars \cite{HaimanLoeb01, MadauRees01} or through the `direct collapse'
of atomic-cooling gas
\cite{BrommLoeb03, Koushiappas+04, Begelman+06, LodatoNatarajan06, SpaansSilk06, Dijkstra+08, Shang+10, Agarwal+12, IO12, Latif+13, Prieto+13, Fernandez+14, TL14, Visbal+14}.
The notion that SMBHs accreted nearly uninterrupted
runs counter to expectations from observations at lower redshifts ($z\lta 2$),
where only a small fraction of SMBHs are undergoing quasar episodes,
which are estimated to last for $1$ to $100 \Myr$
\cite[e.g.][]{Haiman+04, Wang+06, Shankar+09a, Shankar+13}.

In this paper, I show that steady and prolific growth 
of nuclear BHs at $z>6$ can be reconciled with their relative
inactivity at lower redshifts if gas accretion episodes near
the Eddington limit are triggered by major mergers
of the BH's host galaxy or dark matter halo.
Major mergers of galaxies have long been associated
with quasar activity \cite{Sanders+88, BarnesHernquist91, Bahcall+97, KauffmannHaehnelt00, Taniguchi04, Li+07, Hopkins+08, Urrutia+08, Shen09, Treister+10, Treister+12, McGreer+14}.
(Note, however, that luminous accretion can also be triggered by secular
processes \cite[e.g.][]{Grogin+05, Georgakakis+09, Cisternas+11, DraperBallantyne12}.)
Previously, Li et al. \cite[][see \S\ref{sec:mergers}]{Li+07} argued,
by examining the hierarchical growth of an exceptionally massive dark matter halo
a large cosmological simulation,
that host major mergers provide a plausible explanation for the growth of $z\gta 6$
quasar SMBHs. Here, I use a semi-analytic Monte Carlo technique \cite{Zhang+08}
to argue that this is the case generally---i.e. that the most massive dark matter halos
experience a rapid succession of major mergers prior to $z\approx 6$.

The rate of major galaxy mergers
per unit time per dark matter halo evolves extremely rapidly with redshift,
roughly as  $|{\rm d}z/{\rm d}t| \propto (1+z)^{5/2}$,
where the proportionality holds at the large redshifts of interest here.
This is a much steeper dependence than most physical timescales
that plausibly govern the duration of a BH feeding event---for
example, the dynamical time at the virial radius
of dark matter halos scales as $(1+z)^{-3/2}$.
Put another way, we can write the duty cycle of BH growth at any epoch
as 
\beq
f_{\rm duty}=\dot{N}_{\rm trig}~ t_{\rm feed},
\label{eq:fduty}
\eeq
where $\dot{N}_{\rm trig}$ is the frequency of trigger events per BH
and $t_{\rm feed}$ is the typical duration of each feeding episode.
Both of the quantities on the right hand side can depend on factors
such as redshift, BH mass, the masses of the merging galaxies,
and so on---I will return to this point later.
For feeding episodes triggered by major mergers of galaxies,
the trigger rate $\dot{N}_{\rm trig}(z)$
increases so rapidly with redshift that
there is a large range of functions $t_{\rm feed}(z)$ that
would allow for nearly continuous BH growth
($f_{\rm duty}(z)\sim 1$) at high $z$.
I motivate a simple parametrization for $t_{\rm feed}$
for the narrow subpopulation of SMBHs of interest ($z\gta 6$, masses $M\gta 10^9\Msol$),
and delineate the region of parameter space that can explain
their formation at the observed number densities.
An example of a successful growth model is one where major galaxy
mergers trigger fast-feeding episodes lasting for a timescale
comparable to the dynamical time of the host halo.

This paper is organized as follows.
In \S 2, I present additional background
by summarizing general properties of luminous
quasars and by detailing the argument that $z\gta 6$
quasar must have experienced nearly continuous growth if their
accretion was Eddington-limited.
I present in \S 3 results from merger-tree simulations of hierarchical
structure formation, showing the prolific growth histories
of the massive dark matter halos that are likely to host the $z>6$ quasars.
In \S 4, I motivate a specific parametrization of the growth episodes
triggered by galaxy mergers,
and model the durations of such episodes
required to explain the $z\sim 6$ quasar SMBHs
without super-Eddington accretion.
I conclude in \S 5.

Throughout this work, I adopt the cosmological parameters
$h=0.7$, $\Omega_0=0.3$, $\Omega_{\Lambda}=0.7$,
$n_{\rm s}=0.96$ and $\sigma_8=0.83$;
these values are chosen based on
the latest published empirical values \cite{Hinshaw+13, PlanckParams}.
While quantities such as the age of the Universe
and the mass function of dark matter halos
are sensitive to these parameters,
the methods and results presented here are
qualitatively robust.

\section{Luminous Quasars and SMBH Growth}
\label{sec:quasars}

The majority of the mass in SMBHs in the local Universe
appears to have been accumulated during luminous
quasar episodes \cite{Merloni04, Hopkins+06}. 
The brightest quasars have
luminosities on the order of $0.1 - 1$ times the Eddington luminosity
of the SMBH engine \cite[e.g.][]{Kollmeier+06, Netzger+07},
$L_{\rm Edd}(M)=1.3\times  10^{38} (M/{\rm M}_{\odot}) \erg\s^{-1}$.
The luminosity can be expressed in terms of the mass energy
of the accreted fuel and a radiative efficiency factor $\eta$ as
\beq
L=\eta\dot{M}c^2.
\eeq
By comparing the cumulative quasar luminosity density at $z \lta 4$
with the mass density in nuclear SMBHs, 
the cosmic mean value of $\eta$ is found to be $\approx 0.07$
\cite{MerloniHeinz08, Shankar+09a},
in rough agreement with theoretical expectations
of luminous accretion flows \cite[][and refs. therein]{Shapiro05}.
The quasar duty cycle, or the fraction of
time SMBHs spend as quasars, is
$\sim 1\%$ at $z\sim 2$ (where the number density of quasars peaks)
\cite[e.g.][]{Shen+07},
and decreases with redshift
\cite[e.g.][]{Haiman+04, White+08, Shankar+10, Shankar+10b}.

Estimates of quasar lifetimes vary,
but tend to fall in the range $\sim 10^6 - 10^8 \yr$
\cite[e.g.][]{Richstone+98, MartiniWeinberg01, HaimanHui01, Hopkins+09, WL09}.
The radiative and kinetic output from the luminous quasar is thought to act
as a negative feedback, making growth intermittent
\cite[e.g.][]{CO01, diMatteo+05, CO07, Yuan+09, ParkRicotti12}.
Although luminous accretion activity is associated with
major and minor mergers, as well as possibly secular processes,
the detailed mechanism that channels the gas to the nuclear SMBH
remains an open topic of study \cite[e.g.][]{HopkinsQuataert10}.
There is uncertainty as to how much of the SMBH growth occurs
when it is observable as a quasar, as opposed to
other stages---such as
in the midst of a galaxy merger---during which the central SMBH
is heavily obscured. 

For a fixed value of $\eta$, Eddington-limited accretion
implies exponential growth ($\dot{M}\propto M$),
with an $e$-folding timescale
\beq
t_{\rm Edd}=450 \eta ~\Myr,
\eeq
with $t_{\rm Edd}=31\Myr$ for $\eta=0.07$.
For the adopted cosmological parameters,
the age of the Universe at $z=6$ ($z=7$)
is $910$ Myr ($750$ Myr).
If the $z\gta 6$ SMBHs grew from the remnants of the
first generation of stars (Population III or `PopIII' stars)
at $z\gta 30$, then they would have had approximately $t_{\rm avail}\approx 700\Myr$
to grow to $\sim 10^9\Msol$.
(Note that the available growth time $t_{\rm avail}$ is only marginally longer
 if the `seed' BH began to grow at, say, $z=50$.)
 
Nuclear BHs can also grow through hierarchical BH mergers,
but the efficiency of this avenue is limited  \cite{VolRees06}
by the gravitational recoil effect \cite[the momentum imparted 
by asymmetric gravitational-wave emission on
the product of a BH merger, e.g.][]{Bekenstein73, Baker+06b, Campan+07, Lousto+10}---that is, Nature cannot simply throw together a thousand seed BHs
to form a BH a thousand times more massive.
Optimistic estimates suggest that mergers between PopIII remnants
could contribute a factor $X_{\rm merge}\sim 100$ toward assembling
a $M\sim 10^9\Msol$ SMBH
by $z\approx 6-7$ \cite{TH09, TPH12, TLH13}, regardless of whether
the seeds formed and began to grow
in halos with virial temperatures $\sim 400 \K$ or $\sim 2000 \K$.

The mean Eddington ratio required to form a SMBH with mass
$M_{\rm SMBH}$ from a seed BH with mass $M_{\rm seed}$
can be written as \cite[equation reproduced from][]{TL14}
\begin{eqnarray}
f_{\rm Edd}&\approx& \ln \left(\frac{M_{\rm SMBH}}{X_{\rm merge}~M_{\rm seed}}\right)\Big/
\left[\frac{t_{\rm avail}}{(\eta/0.07)\,t_{\rm Edd}}\right]\nonumber\\
&\approx&
\left[0.676+0.045\ln\left(\frac{M_{\rm SMBH}}{3\times 10^9\,\Msol} \frac{30\,\Msol}{M_{\rm seed}}\frac{30}{X_{\rm merge}}\right)
\right]\nonumber\\
&&\qquad \times\left(\frac{\eta}{0.07}\right)\left(\frac{t_{\rm avail}}{700\,\Myr}\right)^{-1}.
\label{eq:fEddPopIII}
\end{eqnarray}
The masses of the $z\gta 6$ quasars could be explained
if they began as PopIII remnants and grew at the Eddington limit
for $\approx 70\%$ of the available time,
or at $\approx 70\%$ of the Eddington limit for the entire available time.

We can repeat the above exercise for the `direct collapse' seed scenario,
in which nuclear BHs form in the gravitational collapse of massive,
atomic-cooling (temperature $T\sim 10^4\K$) gas clouds.
Such an event could occur if the gas has low metallicity
and is inundated by a strong ultraviolet (UV) flux
that photo-dissociates molecular hydrogen and thus prevents
fragmentation of the cloud into stars of ordinary mass
\cite[][additional references in \S\ref{sec:intro}]{OH02, BrommLoeb03}. 
Direct collapse could form BHs with masses $10^4-10^5\Msol$,
but only after redshifts $z\approx 15$ \cite{Dijkstra+08, Agarwal+12, Fernandez+14}
(when the Universe is $\sim 300\Myr$ old)
if the mechanism is requires the prior emergence of powerful
UV sources \cite[see][for a discussion]{Visbal+14}.
In other words, direct collapse seeds are expected to begin
with a head start in mass, but a delayed start in time.
Moreover, because these seeds can only form in rare massive
halos under specific circumstances,
their opportunities to grow via major mergers is limited
(i.e. smaller $X_{\rm merge}$ than a scenario where PopIII seeds form and merge).
We have \cite{TL14}
\begin{eqnarray}
f_{\rm Edd}\approx
&
\left[0.580+0.063\ln\left(\frac{M_{\rm SMBH}}{3\times 10^9\,\Msol} \frac{10^5\,\Msol}{M_{\rm seed}}\frac{3}{X_{\rm merge}}\right)
\right]\nonumber\\
&\qquad \times\left(\frac{\eta}{0.07}\right)\left(\frac{t_{\rm avail}}{500\,\Myr}\right)^{-1}.
\label{eq:fEddUV}
\end{eqnarray}

Equations \ref{eq:fEddPopIII} and \ref{eq:fEddUV}
imply that \textit{both the PopIII and direct collapse seed scenarios
require the $z\gta 6$ quasar SMBHs to have grown nearly continuously}
 (i.e. accreting more than half of the time),
if the growth occurred at rates near the Eddington limit.
This statement is still true even if $M_{\rm seed}\sim 10^4\Msol$
BHs were to have formed as early as $z\sim 30$ \cite[see][for an example of such a scenario]{TL14}.
The requirement of such large duty cycles poses a stark contrast
with the rarity of luminous quasar activity observed at $z\lta 2$.
It is this contrast that I address in this paper.

\section{The Prolific Merger Histories of $z\gta 6$ Quasar Hosts}
\label{sec:mergers}

I begin by discussing the frequency of major merger events
for massive galaxies at redshifts $z\gta 6$. 
The goal here is to answer a simple question:
supposing that a galaxy major merger triggers a feeding episode of the central BH,
how often do the hosts of the $z\gta 6$ quasar SMBHs undergo such triggers?
In other words, I seek to quantify one of the two factors, $\dot{N}_{\rm trig}$ 
on the right hand side of interest in equation \ref{eq:fduty};
I will turn to the other factor, $t_{\rm feed}$, in the next section.

Strictly speaking, throughout this paper I am referring to mergers
of host dark matter halos, not host galaxies
However, at redshifts of interest, the masses of the most
massive halos ($M\sim 10^{13}\Msol$) are
comparable to the largest galaxy masses
\cite[e.g.][]{ReesOstriker77, OA+14}, but not to galaxy clusters. 
Therefore, in this paper I assume
that galaxy counts and halo counts are equivalent,
and use the two terms interchangeably.

\subsection{Major merger histories of the most massive $z\approx 6$ halos}
The number density of $\sim 10^9\Msol$ SMBHs at $z\sim 6$
is $n\sim 1\;{\rm comoving}\Gpc^{-3}$ \cite{Willott+10b},
or of order one or a few in the Millennium Simulation \cite{Springel+05} volume.
The masses of dark matter halos with comparable abundances
are $\approx 10^{13}\Msol$, and the naive expectation is that these SMBHs
are hosted by the most massive class of halos \cite[see, however][]{Fanidakis+13}.
Earlier, Li et al. (2007) \cite{Li+07} examined a particularly massive dark matter halo
in a cosmological $N$-body simulation and noted that it
experienced seven major mergers
(which they defined as mergers with progenitor mass ratio $\xi\ge 1/5$)
in rapid succession between $z=14$ and $z=6$. They showed that this
prolific merger history provides a plausible explanation for the growth of a $z\gta 6$,
$M\gta 10^9\Msol$ SMBH without super-Eddington accretion.
Angulo et al. (2012) \cite{Angulo+12} noted that the most massive
$z\approx 6$ halos ($\approx 10^{13}\Msol$) doubled their masses
in the preceding 100 Myr.

\begin{figure}[t]
\begin{center}
\includegraphics[width=\textwidth]{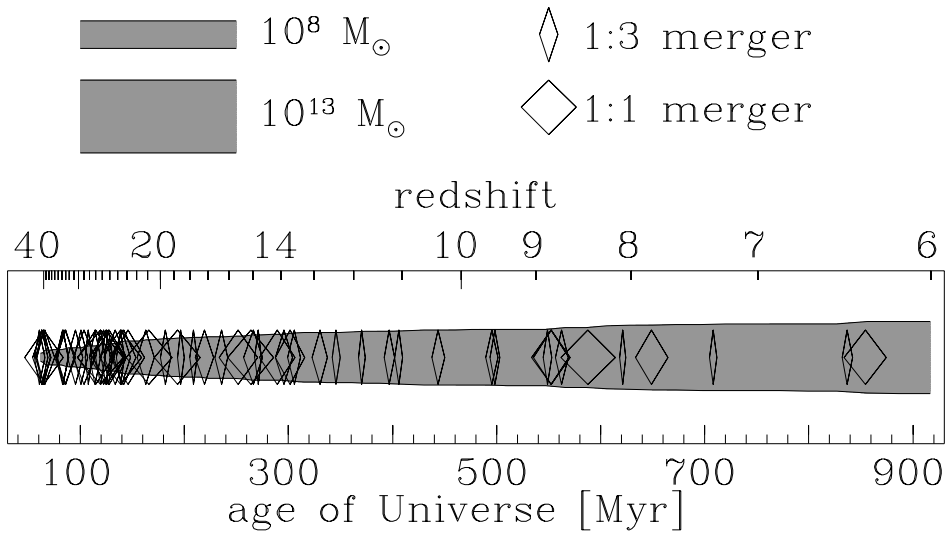}
\end{center}
\caption{\label{fig:merge1} \footnotesize  
The growth history of one DM halo with $M(z=6)\approx 10^{13}\Msol$.
The thickness of the gray shaded region is proportional to $\log[M_{\rm trunk}/(10^5\Msol)]$,
where $M_{\rm trunk}$ is the most massive progenitor halo (the ``trunk'' of the merger tree).
The diamonds denote significant merger events (progenitor masses more equal than 1:9),
with the center of the diamond marking the time of the merger and
the width-to-height ratio equal to the progenitor mass ratio.}
\vspace{-1\baselineskip}
\end{figure}

I present that the explanation provided by Li et al., if qualitatively correct,
is likely to hold quite generally for any comparable volume in the Universe.
Figure \ref{fig:merge1} shows a graphical representation of the
merger history of a dark matter halo that reaches a mass of
$M\approx 10^{13}\Msol$ at $z=6$.
The merger history is generated using a Monte Carlo merger tree algorithm
whose underlying mathematical formalism provides excellent matches
to cosmological $N$-body simulations,
especially for large halo masses \cite[e.g.][]{Reed+07},
and whose numerical implementation has been shown to be highly
accurate \cite{Zhang+08}.

The horizontal axis of Figure \ref{fig:merge1} shows the age of the Universe,
in linear scale to emphasize the rapid hierarchical growth of the halo.
The width of the shaded region represents the mass of the most massive progenitor
of the halo (i.e. the `trunk' of the merger tree) at any given time;
the width of the region is proportional to $\log [M_{\rm halo}/(10^5\Msol)]$.
The diamonds mark merger events (down to mass ratios of $\xi=1/9$),
with the center of the diamond indicating the time of the merger
and the ratio of the axes showing the mass ratio of the merging
progenitors.

Between $z=16$ and $z=6.3$, this particular halo experiences
10 mergers with mass ratios $\xi >1/3$, 
16 mergers with $\xi >1/5$, and 
10 additional mergers with $1/5>\xi>1/9$.
The physical time between mergers is much shorter at higher redshifts,
showing a steep evolution in the merger rate with redshift (more on this point shortly).

\begin{figure}[t]
\begin{center}
\includegraphics[width=4.7in]{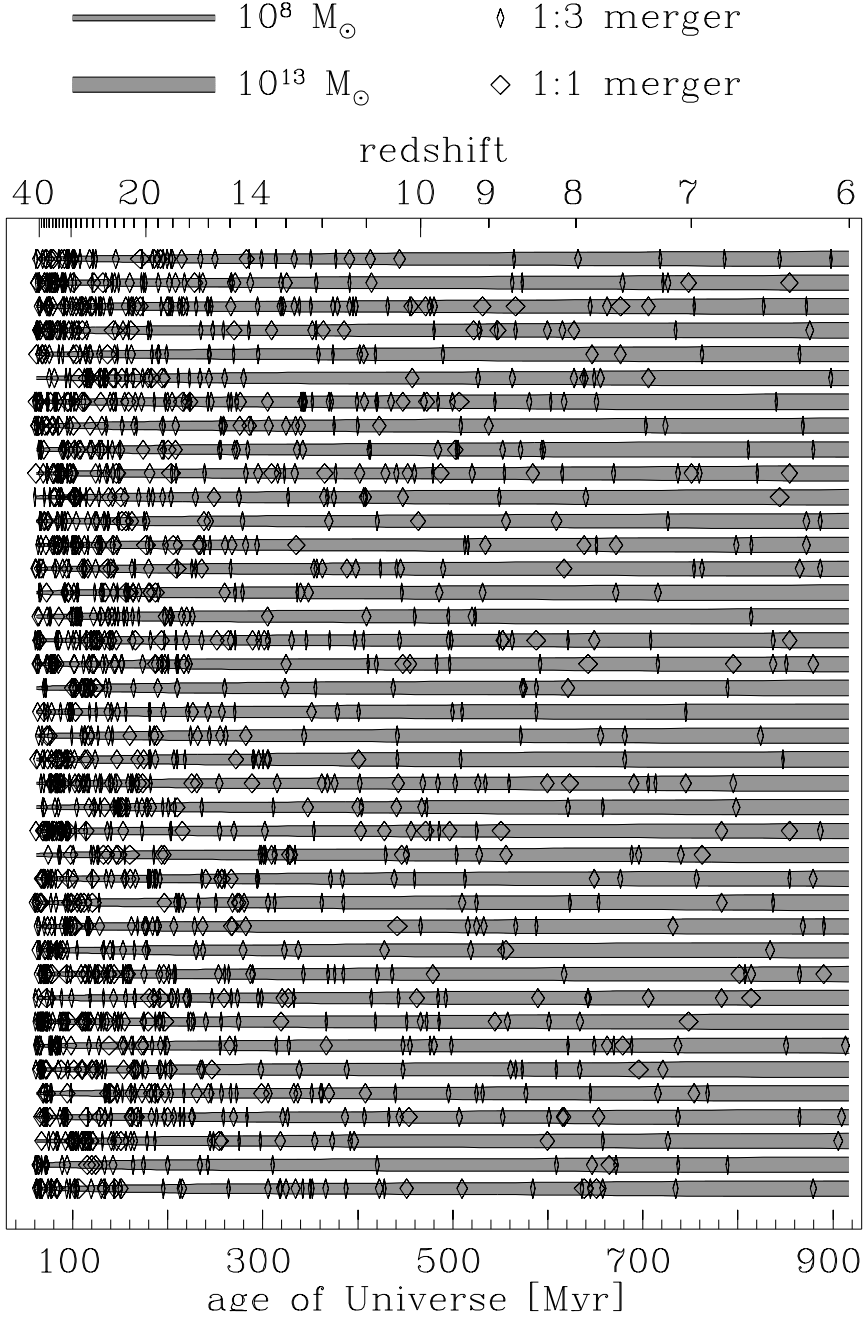}
\end{center}
\caption{\label{fig:merge40} \footnotesize  
The same as Fig.1, except that this figure shows the growth histories of 
40 simulated DM halos with $M(z=6)\approx 10^{13}\Msol$.
The thickness of each gray region is proportional to $\log[M_{\rm trunk}/(10^5\Msol)]$.
The diamonds mark merger events with progenitor masses more equal than 1:9,
with the width-to-height ratio equal to the progenitor mass ratio.
The ``trunk'' of each merger tree has a mass of $\sim 10^6\Msol$ at $z=40$.
}
\vspace{-1\baselineskip}
\end{figure}

Figure \ref{fig:merge40} presents the same information,
but for 40 simulated dark matter halos, all with $M(z=6)>10^{12.9}\Msol$.
The merger tree sample is the same as the one used in Tanaka \& Li (2014) \cite{TL14}.
It is readily apparent from the figure that the points made in the previous paragraph
hold generally for this mass class of halos.
All of these halos undergo $N_{\rm m}\gta 10$ major merger
events (regardless of whether one defines a major merger
with a minimum mass ratio of 1/3 or 1/5) between $z\approx 15$
and $z\approx 6$, a span of $\sim 0.65 \Gyr$.
By contrast, the typical massive galaxy experiences
$\sim 1$ major merger between $z\sim 3$ and $z=0$ \cite[e.g.][]{Man+12},
a span of over $11\Gyr$.

\subsection{The redshift evolution of the major merger rate}

\begin{figure}[t]
\begin{center}
\includegraphics[width=\textwidth]{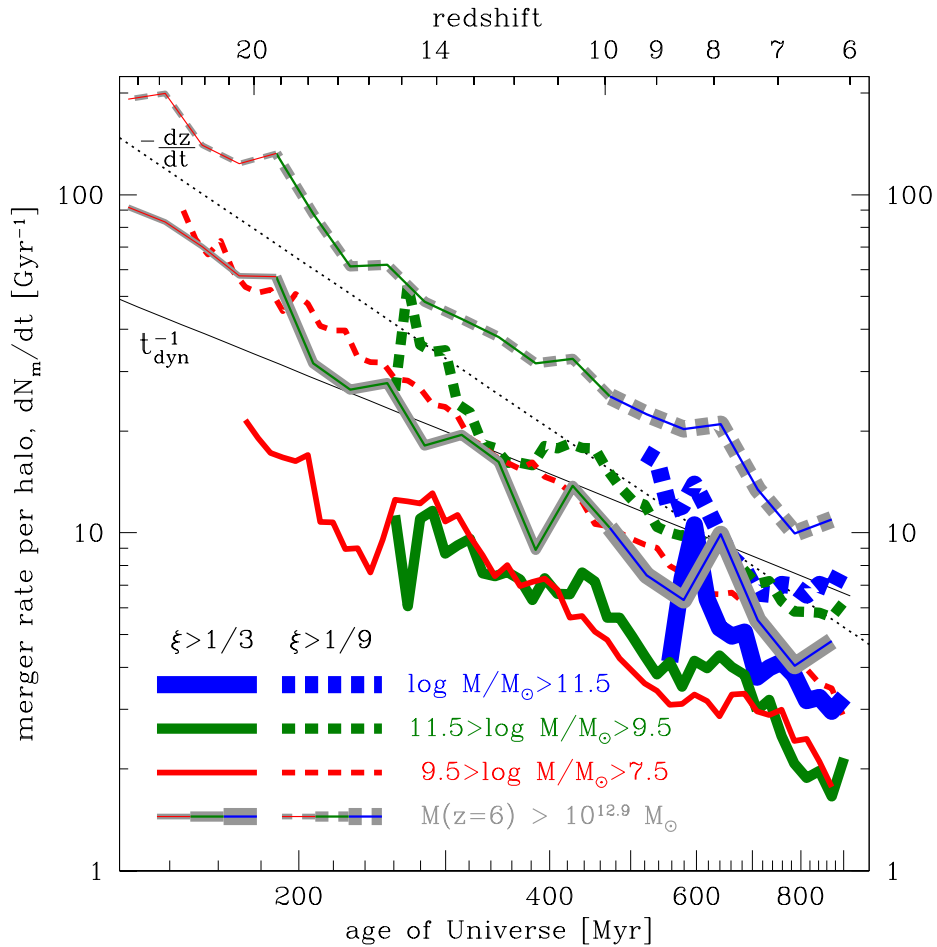}
\end{center}
\caption{\label{fig:mergerates} \footnotesize  
The mean merger rates ${\rm d}N_{\rm m}/{\rm d}t$ (per unit time per halo),
of DM halos in the merger tree simulation.
The thick blue lines denote mergers for halos with $M>10^{11.5}\Msol$,
the green lines for halos with $10^{11.5}\Msol>M>10^{9.5}\Msol$,
and the thin red lines for halos with $10^{9.5}\Msol>M>10^{7.5}\Msol$.
The solid and dashed lines show mergers with progenitor mass ratios $\xi>1/3$
and $\xi>1/9$, respectively.
The grey curves show the mean merger rates for the main (`trunk') progenitor
for halos that have $M>10^{12.9}\Msol$ at $z=6$
(the same 40 halos as shown in the previous figure),
with the line thickness and color accents indicating when the mean mass
of the trunk lies within the three bins described above.
For reference, I have also plotted the quantities 
$t_{\rm dyn}^{-1}\propto (1+z)^{3/2}$ (solid black line)
and
$|{\rm d}z/{\rm d}t|\propto (1+z)^{5/2}$ (dotted black line).
}
\vspace{-1\baselineskip}
\end{figure}

To further emphasize the fact that the prolific merger history found
by Li et al. is generic for all halos of similar mass at similar redshift,
and in an effort to quantify this trend,
I show in Figure \ref{fig:mergerates} the mean merger rates 
${\rm d}N_{\rm m}/{\rm d}t$ (per unit time per halo) of massive halos.
The halo sample includes 40 halos with $\log_{10}[M(z=6)/{\rm M}_{\odot}]>12.9$,
100 halos with $12.9 > \log_{10}[M(z=6)/{\rm M}_{\odot}]>12.5$,
and 100 halos in each mass bin with $12.5 > \log_{10}[M(z=6)/{\rm M}_{\odot}]>12.0$,
$12.0 > \log_{10}[M(z=6)/{\rm M}_{\odot}]>11.5$
etc., down to $8.5 > \log_{10}[M(z=6)/{\rm M}_{\odot}]>8.0$.
The counting statistics for each mass bin are scaled up to match the abundances
expected in a 50 comoving Gpc$^3$ volume \cite[`halo cloning'; see, e.g.][]{TH09, TL14}.
This method is effective for examining the assembly histories of the 
most massive DM halos, which are sparsely sampled by even
the largest cosmological $N$-body simulations;
on the other hand, it offers much weaker statistical power
for all but the most massive objects (i.e. the lower-mass bins are
represented by a hundred halos here, but represented in the
thousands in large $N$-body simulations).

In Figure \ref{fig:mergerates},
the thick blue, medium green, and thin red curves show
merger rates of dark matter halos whose (post-merger) masses
are $\log_{10}[M(z)/{\rm M}_{\odot}]>11.5$,
$11.5>\log_{10}[M(z)/{\rm M}_{\odot}]>9.5$
and $9.5>\log_{10}[M(z)/{\rm M}_{\odot}]>7.5$,
respectively.
The solid curves show the rate of mergers whose progenitor mass
ratios are greater than $\xi>1/3$, and the dashed lines
show the rate of mergers with $\xi>1/9$.

The grey lines show the mean merger rate in the main progenitor
(the merger tree `trunk') of the most massive halos that have
$M>10^{12.9}\Msol$ at $z=6$. This is the same sample of 40 halos
whose merger histories are graphically represented in Figure \ref{fig:merge40}.
The thickness of the grey curves and their colored accents
denote where the mean mass for this sample lies within
the mass bins described above (i.e. thick blue denotes where
$\log_{10}[\langle M(z)\rangle /{\rm M}_{\odot}]>11.5$, and so on).
Note that whereas the blue, green and red curves show the mean
merger rates for halos in the entire simulation sample,
the grey curves show the merger rates only for the main progenitors
of halos that end up with $M(z=6)>10^{12.9}\Msol$.

Figure \ref{fig:mergerates} shows some noteworthy trends,
most of which have been noted in previous studies utilizing
semi-analytic methods \cite[e.g.][]{Benson+05, NeisteinDekel08}
and $N$-body simulations \cite[e.g.][]{Genel+09, Fakhouri+10}.
First, more massive halos have somewhat higher merger rates.
Second, mergers of ratios $\xi >1/3$ are almost as common
as mergers of $\xi>1/9$.
This piece of information is useful because the definition
of what mass ratios constitute a `major' merger is arbitrary;
previous works have used $\xi>1/4$, $\xi>1/5$, and so on.
This figure suggests that the choice of a minimum $\xi$ value
for defining a major merger
(i) does not qualitatively affect the (steep) redshift evolution
of the inferred major merger rate,
and (ii) affects it quantitatively only up to a factor of order unity.
(Note that there are other uncertainties in defining and interpreting
merger rates of halos and galaxies \cite{Hopkins+10}).
Third, and most importantly for the topic of this paper and as suggested
by the previous figures, the major merger rate per galaxy 
evolves very rapidly with redshift.
Across the range of halo masses considered here, the rate at $15 \lta z\lta 20$ is higher by
an order of magnitude than the rate at $6 \lta z\lta 8$,
which is in turn $\sim 10$ times greater than
the predicted major merger rates of quasar hosts
at $z\approx 2$ \cite[][see below]{Fakhouri+10}.
Finally,  the cosmic mean merger rate for
\textit{halos in a fixed mass range at all redshifts} (colored curves)
and the mean merger rate for the main progenitors of the
$M(z=6)\approx 10^{13}\Msol$ halos \textit{across five orders of magnitude
of mass growth} (grey curves) exhibit similar redshift evolution.

For comparison, I have plotted the quantity $|{\rm d}z/{\rm d}t|$,
or the rate at which the Universe ages per unit redshift, 
as a dotted black curve. At the redshift values of interest,
${\rm d}z/{\rm d}t$ scales approximately as $\propto (1+z)^{5/2}$.
The evolution of the halo merger rate roughly follows
this power law, consistent with theoretical expectations
and confirming the fidelity of the merger trees.
This simply reflects the fact that the theoretical
merger rate per redshift ${\rm d}N_{\rm m}/{\rm d}z$ depends weakly on $z$.

These semi-analytic results can be compared to
those of Fakhouri et al. (2010) \cite{Fakhouri+10},
who investigated the halo merger rates in the two Millennium Simulations
\cite{Springel+05, BoylanKolchin+09}.
Although the cosmological parameters adopted here
differ slightly from those in those simulations, 
the two sets of results are broadly consistent with each other:
those authors also found major merger rates of ${\rm d}N_{\rm m}/{\rm d}t \gta 1 \Gyr^{-1}$
for massive halos at $z\gta 6$, and that ${\rm d}N_{\rm m}/{\rm d}t$
scaled roughly as $\propto {\rm d}z/{\rm d}t\times (1+z)^{0.1}\propto (1+z)^{2.6}$.
The reader may wish to juxtapose Figure \ref{fig:mergerates} 
in this work to the right-hand panel of Figure 3 in Fakhouri et al.
Note that whereas that study focused mostly on halo mergers
at $z\ltsim 7$, here I'm interested in the range $6\le z \lta 40$.

\section{SMBH Feeding Times}
Having quantified the host merger rates, the rate of trigger events
in equation \ref{eq:fduty}, I now turn to the effective duration
of the SMBH `feeding time.'

It is worth re-emphasizing that the exact manner in which galaxy mergers
deliver gas to the central BH(s) is not fully understood
\cite[cf.][]{HopkinsQuataert10, HopkinsQuataert11}.
Assuming constant growth at fixed Eddington rate is (predictably)
problematic \cite{TH09}.
While numerous studies---semi-analytic models, as well as 
simulations that employ sub-grid prescriptions for SMBH growth---have
used Bondi-Hoyle accretion,
this prescription is known to break down in the presence of
radiative feedback from the BH \cite{ParkRicotti12},
angular momentum \cite{Li+13},
inhomogeneous gas cooling and dynamics \cite{Hobbs+12, Gaspari+13},
advection \cite{NY94, BB99}, thermal conduction \cite{TM06, JohnsonQuataert07}, etc.
Indeed, many examples of luminous BH activity do not appear to
be represented by Bondi accretion \cite[e.g.][]{Kuo+14}.
Prescriptions where SMBH growth is coupled to
the baryonic properties of the host can be successful
with specific model prescriptions
\cite[e.g.][]{VHM03, Bromley+04}, as are some models
with local \cite[e.g.][]{diMatteo+05,MerloniHeinz08}, 
and global \cite{TPH12} regulatory feedback.

In discussing the host merger-triggered feeding times,
it is important to clarify two points.
First, the growth episode is not a step-function event \cite[e.g.][]{Hopkins+09},
but rather a continuous process during which the SMBH accretion rate
varies with time. Here, I refer to as the `feeding time' the effective
period over which the average accretion rate is equal to Eddington.
Second, I distinguish the terms `feeding time' and `quasar lifetime'
to emphasize that the two need not be the same.
Although quasar episodes are associated with the final stages
of SMBH growth via rapid gas accretion,
at high redshifts the growth could be preferentially more obscured
due to greater gas column densities---particularly so,
if obscuration by host mergers are commonplace.
Depending on the host morphology, as well as the line of sight, timing
and wavelength of the observation,
a given growth episode may or may not be classified as a quasar.

In general, the feeding time should depend on a number
of variables: the total mass and mass ratio of the merging host,
the redshift, the mass and spin of each nuclear BH,
the gas metallicity, 
the geometry of a given host merger and
the angular momentum of the gas,
and perhaps the dynamics of the nuclear BHs
as they form a binary and evolve.
However, in this particular instance, we're concerned
with a very specific subpopulation of massive dark matter halos
that share similar growth histories.
All of these halos were selected to have similar masses at $z=6$,
and their mass growth histories $M(z)$ are highly uniform,
with the masses of their main progenitors typically only varying by
a factor of a few out to $z\sim 40$; see Fig. \ref{fig:merge40}.
(Note that this trend does not extend to lower redshifts;
the most massive halos at $z\sim 6$ do not necessarily
grow into the most massive halos at $z\lta 2$ \cite[e.g.][]{Angulo+12}.)
I will also restrict the following analysis to major mergers
with mass ratios $\xi \ge 1/5$.

In addition, these halos always have
masses well above the cosmological Jeans (filtering) scales,
even in hypothetical scenarios where the intergalactic medium
is heated prolifically by early mini-quasar activity \cite{TPH12}.
Thus, at any given redshift, these halos are the least sensitive (compared
to lower-mass halos) to spatial fluctuations in local radiative backgrounds.

I consider a feeding time of the form
\beq
t_{\rm feed}(M_{\rm halo}, M_{\rm BH}, z, \xi, ...)
\sim \left\langle t_{\rm feed} (z=6) \right\rangle \left(\frac{1+z}{7}\right)^A.
\label{eq:tfeed}
\eeq
That is, because in this particular instance the variations in
$M_{\rm halo}(z)$, the merger rates, and the mass ratio $\xi$
are small,
I suppose that the duration of the SMBH feeding episode
can be characterized by a characteristic mean value
$\left\langle t_{\rm feed} (z) \right\rangle$
that follows a redshift evolution $(1+z)^A$.
I assume that at any given redshift, the feeding episode duration
can be characterized by a log-normal distribution with scatter $B$.

Many studies have sought to empirically estimate
quasar lifetimes, with large uncertainties.
The best constraints come from data at $z\lta 2$;
the limited redshift range, and the fact that
the overall quasar sample contains
a wide variety of host galaxies and SMBH masses,
makes it difficult to evaluate how SMBH growth depends on the redshift
and the host properties.
For example, Wyithe \& Loeb (2009) \cite{WL09} suggest that
the quasar lifetime scales with the dynamical time
of the host dark matter halo, 
$t_{\rm dyn} \approx 230 \; [(1+z)/7]^{-3/2} \Myr$ \cite[e.g.][]{BarkanaLoeb01}.
For reference, I have plotted $t_{\rm dyn}^{-1}\propto (1+z)^{3/2}$
alongside the halo merger rates in Figure \ref{fig:mergerates} (solid gray line).

I perform a parameter space survey to quantify what combinations
of feeding time parameters---normalization $\left\langle t_{\rm feed} (z=6) \right\rangle$,
redshift evolution slope $A$, and scatter $B$---can explain 
the observed population of $z\sim 6$, $M\gta 10^9\Msol$ quasars via
host merger-triggered, Eddington-limited growth.
I assume that a nuclear BH is in place in each massive halo,
and that following a major merger $\xi \ge 1/5$ the BH
accretes for $t_{\rm feed}$, drawn randomly
out of the $z$-dependent log-normal distribution,
over which time the mean accretion rate is Eddington.
I take a conservative model, in which a subsequent major merger
during $t_{\rm feed}$ does not extend the feeding episode
(i.e. rapid mergers cannot counteract feedback by `blowout');
an alternative model would be one where the BH accretes
for $t_{\rm feed}$ since the last major merger.
This exercise is repeated 100 times (to statistically average different possible
Monte Carlo realizations of $t_{\rm feed}$)
for each of the main progenitors of
the sample of halos with $M(z=6)\ge 10^{12}\Msol$.

Molecular-cooling halos that have just formed PopIII stars
will have shallow potentials, and may be more susceptible
to negative feedback from BH activity \cite{Alvarez+09, Milos+09}.
Therefore, I only allow BHs to grow if their halos are atomic-cooling 
(virial temperatures $\sim 10^4\K$),
which may be a crucial threshold that allows for dense cooling flows
to carry gas to the central BH \cite{WiseAbel07, Greif+08, DiMatteo+12}.
Note that this sample of particularly massive halos
becomes atomic-cooling at $z\gta 30$,
and exist at the same abundances
as the $z\gta 6$ quasars ($\sim \Gpc^{-3}$) as early as $z\sim 40$ \cite[see][Fig. 2]{TL14}.
This is much earlier than the typical redshift for PopIII formation ($z\approx 20$)
or proto-galaxy formation ($z\gta 10$).
This implies that if the first `monster' SMBHs grew from PopIII seeds,
then the numerical simulations focusing on typical halos at $z\sim 10-20$
may not be representative of their cradles. 

\begin{figure}[t]
\begin{center}
\includegraphics[width=\textwidth]{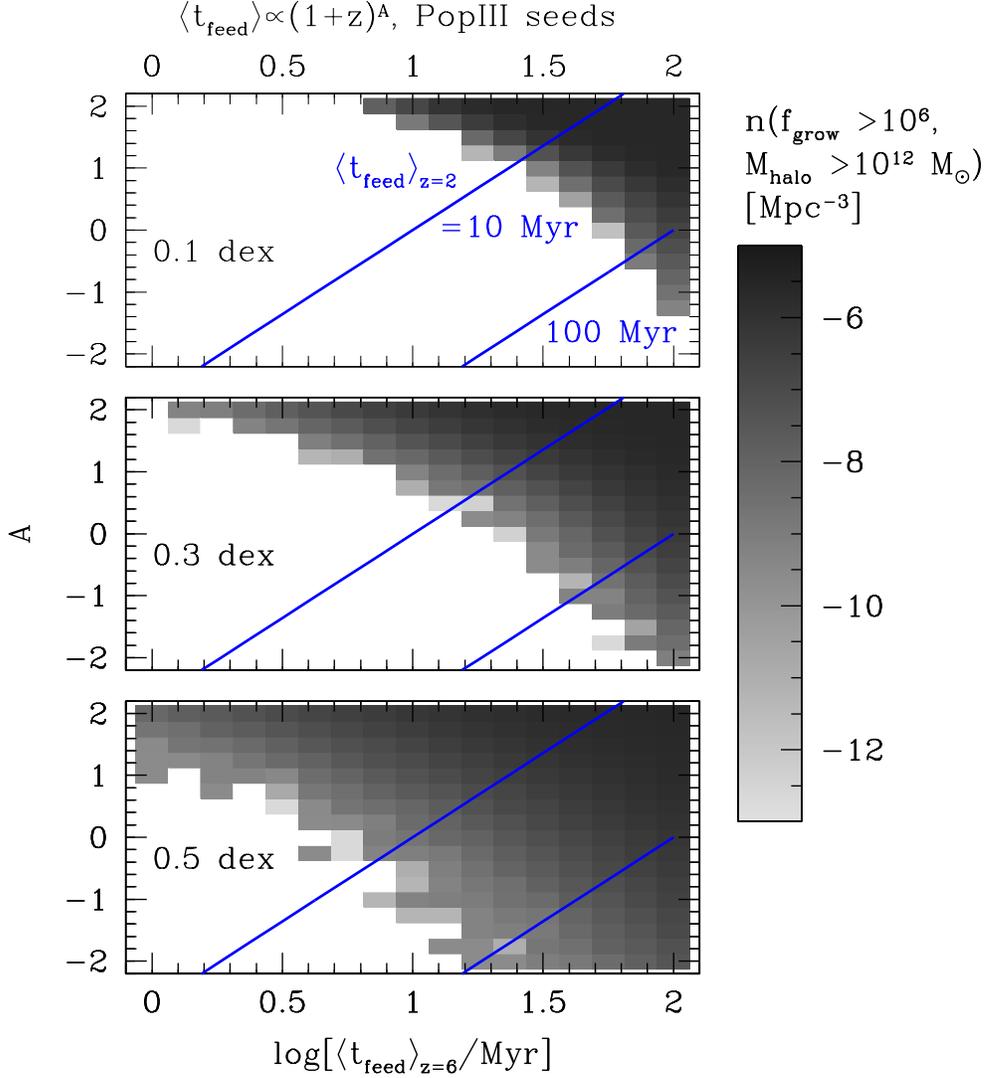}
\end{center}
\caption{\label{fig:npop3} \footnotesize  
The number density of BHs in halos with $M_{\rm halo}(z=6)\ge 10^{12}\Msol$,
and that have grown by a factor $f_{\rm grow}\ge 10^6$
between when the main progenitors of host halo becomes atomic-cooling
(virial temperature $\ge 10^4\K$, $z\gsim 30$) and $z=6$.
These criteria are representative of the growth requirements from a Population III
BH seed (see text).
The different grids show outcomes for different combinations of parameters
for the feeding time:
the normalization at $z=6$
and the redshift dependence exponent $A$ (where $\langle t_{\rm feed}(z)\rangle \propto (1+z)^A$).
The three panels show results for different assumptions of
log-normal scatter in the feeding times, from top to bottom, 0.1, 0.3 and 0.5 dex.
White grids show models that have
no SMBHs at $z=6$, and dark gray grids show those that exceed
the observed quasar number density ($n\sim 10^{-9}\Mpc^{-3}$).
}
\vspace{-1\baselineskip}
\end{figure}

The collected output is the number density of host halos
whose central BHs would have grown to $\sim 10^9\Msol$ by $z=6$.
In Figure \ref{fig:npop3}, I plot this quantity for 
population synthesis realizations resulting from
different combinations of the three model parameter values.
I evaluate the number density of halos whose BHs have grown
by a factor $f_{\rm grow}\ge 10^6$ from when the main progenitor
of the host halo is atomic-cooling to $z=6$.
That is, the seed BH in the main progenitor
must grow by a factor of $\ge 10^6$ via
host merger-triggered gas accretion episodes, 
and acquire $1000\Msol$ from a combination of the initial mass
and BH mergers (e.g. by having $100\Msol$ at formation
and growing by a factor $10$ via BH mergers).
The white grids in the figure represent model parameter
combinations that do not produce any such quasar SMBHs.
The dark grey and black grids show models that overproduce
massive BHs at $z=6$.
A model may be said to be viable if the predicted number density
of $z\sim 6$, $M\gta 10^9\Msol$  SMBHs matches the
observed estimate of $n\sim 10^{-9}\Mpc^{-3}$ \cite{Willott+10b}.

The blue solid lines in Figure \ref{fig:npop3} show, for reference, the combination
of parameters where $\langle t_{\rm feed}(z=2)\rangle$
would be $10 \Myr$ and $100 \Myr$, representing the
approximate quasar lifetimes derived from observations at $z\lta 2$.
These blue lines are meant only as rough guides. Again,
(i) quasar lifetimes are not (necessarily) BH feeding times,
and (ii) the halos studied here belong to a very narrow subset
of the most massive $z>6$ halos and have quantitatively similar
mass growth histories, and may be quite different from hosts of low-redshift quasars.

\begin{figure}[t]
\begin{center}
\includegraphics[width=\textwidth]{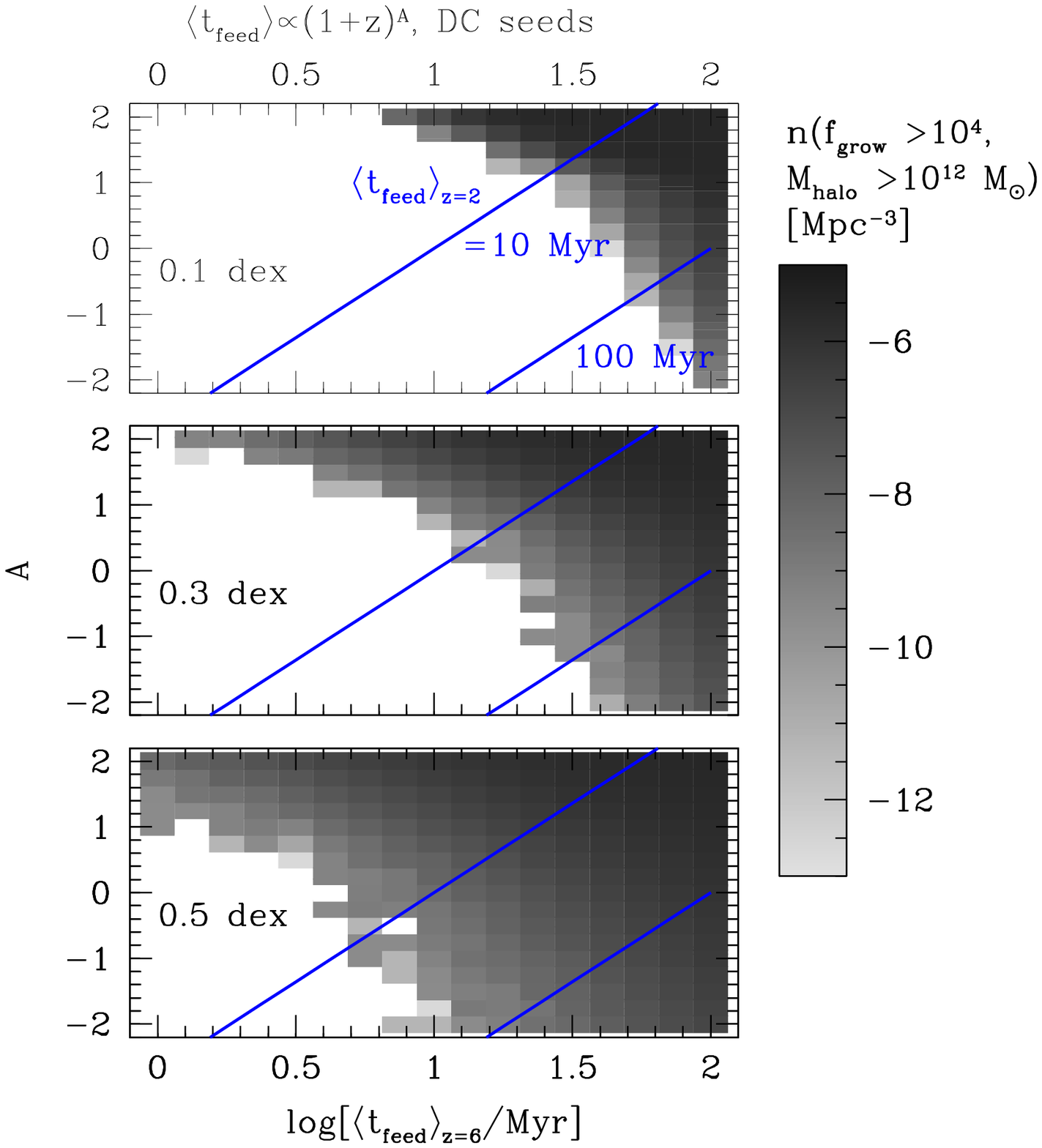}
\end{center}
\caption{\label{fig:ndc} \footnotesize  
Same as the previous figure, but here I plot the number density
of BHs that have grown by a factor 
$f_{\rm grow}\ge 10^4$
between $z=15$ and $z=6$, representative of the growth requirements
for UV-aided direct collapse seed models.
}
\vspace{-1\baselineskip}
\end{figure}

I repeat the above parameter survey for direct-collapse seed models,
which may form by $z\approx 15$ and have initial masses as large
as $M\sim 10^5\Msol$ [see \S\ref{sec:quasars}].
Whereas for Population III seed models I considered BHs
that have grown by $f_{\rm grow}\ge 10^6$ between when the
host becomes atomic-cooling and $z=6$, here I consider
BHs that have grown by $f_{\rm grow}\ge 10^4$ between $15\ge z \ge 6$.\footnote{
Note that the analysis does not require the seed BH to form in the
main progenitor; it can form in a UV-inundated satellite halo
in the vicinity of the `trunk' halo, which serves as the UV source
to aid direct collapse \cite{Dijkstra+08}, then subsequently fall into the more massive halo.}
The results are plotted in Figure \ref{fig:ndc}.
The parameter requirements do not differ very much from the PopIII
case, which is not surprising given that similar mean Eddington ratios
are required for both the PopIII and direct collapse families of models
[see equations \ref{eq:fEddPopIII} and \ref{eq:fEddUV}].
Roughly speaking, both models require BHs of masses $\sim 10^5\Msol$
to be in place by $z\sim 12-15$, and grow by a factor $\sim 10^4$ by $z\approx 6$.

Figures \ref{fig:npop3} and \ref{fig:ndc} show that the
$z\gta 6$ quasar population could be explained,
for both Population III and direct-collapse seed models,
by host merger-triggered accretion near the Eddington limit if feeding episodes
last for $\left\langle t_{\rm feed} (z=6) \right\rangle\sim 30-100\Myr$
and if the slope of the redshift evolution is $A\gta -2$.
Large values of the scatter $B\gta 0.3$  in the distribution of feeding times
can enhance the number of particularly massive BHs
(i.e. if it is more common for major mergers to trigger long
feeding episodes).

\section{Conclusions}
\label{sec:concl}

I have investigated the merger histories of dark matter halos
at $z\gta 6$, focusing on halos that are massive enough ($M(z=6)\ge 10^{12}\Msol$)
to plausibly host the $z\gta 6$ quasars.
Below is a summary of the findings.

\begin{enumerate}
\item The mean major merger rate of the main progenitors of
the $M(z=6)\approx 10^{13}\Msol$ halos
is approximately equal to $|{\rm d}z/{\rm d}t|\propto (1+z)^{5/2}$, 
all the way up to redshifts $z\sim 40$.
This approximation is valid within a factor of two,
whether one defines a major merger by a mass ratio threshold
$\xi>1/9$ or by $\xi>1/3$.
This result is consistent with expectations from semianalytic
theory and results from large cosmological $N$-body simulations.
\item The steep evolution of the merger rate ${\rm d}N_{\rm m}/{\rm d}t$
directly implies that for a wide range of physical BH feeding mechanisms
with duration $t_{\rm feed}$,
the duty cycle $f_{\rm duty}={\rm d}N_{\rm m}/{\rm d}t\times t_{\rm feed}$
increases with redshift.
While $t_{\rm feed}$ can in general depend on a myriad factors,
I took advantage of the fact that the halos of interest here share
closely similar assembly histories to conjecture that the feeding times
for these halos can be characterized as a function of redshift
with a reasonably small scatter.
The formation of the $z\gta 6$ quasar SMBHs can be explained
without super-Eddington accretion, for both PopIII and direct collapse
seed models, if $t_{\rm feed}$ is greater than several $10 \Myr$ at $z\approx 6$
and scales with redshift as $(1+z)^A$ with $A\gta -1.5$, for scatter in 
$t_{\rm feed}$ of $\sim 0.3$ dex.
This parameter space includes $t_{\rm feed}\lta t_{\rm dyn}(z)= 230 \; [(1+z)/7]^{-3/2} \Myr$.
This finding suggests that the SMBH growth scenario suggested by Li et al. \cite{Li+07}
may be viable for a plurality of all dark matter halos at this mass scale that host a nuclear BH.
\item The main progenitors of the $z\gta 6$ quasar hosts
become molecular-cooling (atomic-cooling)
very early, at $z\gta 40$ ($z\gta 30$),
significantly earlier than halos in detailed cosmological
simulations that focus on the formation of typical
first stars and galaxies, at $z\gta20$ ($z\gta 10$).
Such simulations may not be well-suited for studying the 
the evolution of the $z\gta 6$ quasar SMBHs,
as they would underestimate the effects of halo growth and mergers,
especially for PopIII seed models.
\end{enumerate}

When considering the origins of the $z\gta 6$ quasar SMBHs,
it is important to keep in mind that the observed properties of $z\gta6$ \textit{galaxies}
are remarkably similar to what is found at later cosmological epochs:
their masses are comparable to (but somewhat lower than)
those of the most massive SMBHs and galaxies in the local Universe;
the quasar spectra and metallicities appear identical to what is found
at lower redshifts \cite{Juarez+09, Mortlock+11};
they exhibit strong star formation \cite{Wang+13} and winds \cite{Maiolino+12}
associated with post-merger quasar activity.
In a similar vein, massive galaxies at $z\sim 7-9$
appear to already have metal-enriched stellar populations
with ages of $\sim 100 \Myr$ \cite[e.g.][]{Dunlop+13,Labbe+13}.

The masses of both the $z\gta 6$ quasar SMBHs
and their hosts are comparable to the largest masses of
SMBHs and galaxies found at lower redshifts.
Indeed, the simple expectation from the theory of hierarchical structure formation
is that the abundance of galaxies approaching the mass ceiling of galaxies
\cite{ReesOstriker77} reaches $\sim \Gpc^{-3}$ by $z\gta 6$.
That is, the observation of SMBHs at the highest mass scales at $z\gta 6$
is coincident with the emergence of the most massive class of galaxies---expected
from theory, and observed to be fully evolved.
It also follows from theory that 
after their relatively early arrival on the cosmic stage,
the population of massive galaxies should evolve
increasingly slowly as the Universe ages---at
the largest masses due to the inhibition of gas cooling,
and in general due to the precipitous drop in the frequency of major mergers 
(the triggers of starbursts and SMBH growth) toward low redshifts.
This is qualitatively consistent with observed trends in the emergence and growth
of the most massive SMBHs and galaxies throughout cosmic time
\cite{Merloni04, Collins+09}.

One then wonders: aside from their early formation, 
what is extraordinary about the $z\gta 6$ quasars and their hosts?
That their hosts are more compact, gas-rich and rapidly merging
despite having similar masses to their low-redshift counterparts
may result in their BHs having somewhat larger Eddington ratios \cite{Willott+10b}
and being heavily obscured for large periods of time
(in addition to their being possibly obscured at birth \cite{Begelman+06}).
The latter possibility could negatively affect the detectability
of $z>7$ quasars by the James Webb Space Telescope
and Athena+ at rest-frame UV and X-ray frequencies, respectively,
while making large contributions to the cosmic infrared background
\cite[see, e.g.][]{Yue+13b, Helgason+14}.
The correlations between SMBHs and galaxy properties
could differ from what is found in the local Universe \cite[e.g.][]{Wang+13},
depending on the details of the seeding and growth mechanisms \cite{VN09, Agarwal+13}.

The formation and evolution of low-mass galaxies and their nuclear BHs
may have proceeded quite differently at high redshifts, as gas cooling in small (proto-) galaxies
is known to be more sensitive to the temperature and ionization state of the inter- and circumgalactic media.
The transition of the IGM from neutral $<100\K$ gas at $z\sim 20$,
to ionized gas with temperatures comparable to protogalactic virial temperatures and
the atomic-cooling threshold at $z\sim 7$,
would have affected star formation and BH growth in low-mass galaxies.
It is interesting to note that the hosts of the $z\gta 6$ SMBHs
were likely to be the least affected by this upheaval of
the intergalactic environment \cite[e.g.][]{TPH12}.

A minimalistic, zeroth-order ansatz would be that
once their immediate environments are ionized
and provided that they are above the cosmological Jeans (filtering) mass scale,
the formation and evolution of galaxies---SMBHs, metallicities, winds and all---are
driven primarily by the gravitational environment of their dark matter halos, at high and low redshift;
that while the earliest galaxies and SMBHs at the largest mass scales arise
and evolve rapidly, they do so without processes
that are either rare or absent (e.g. highly super-Eddington accretion)
in galaxies of similar mass at lower redshift.
Or, we can turn this ansatz into a question:
Is there a redshift above which the evolution of massive
galaxies and their SMBHs is qualitatively different from what is observed at $z\lta 2$?
Just as Turner (1991) \cite{Turner91} noted for $z\sim 4$ quasars,
the formation of the $z\gta 6$ quasars could 
be explained within the confines of `conventional cosmic structure formation.'
\\

\section*{Acknowledgments}

I am grateful to Miao Li for graciously allowing me to re-use
merger tree data from an earlier collaboration.
I thank Eugene Churazov for stimulating discussions,
and Zolt\'an Haiman for encouraging conversations
and comments on the manuscript.\\

\bibliographystyle{unsrt}

\end{document}